\documentclass[letterpaper, 10 pt, conference]{ieeeconf}
\usepackage{amsmath}

\let\labelindent\relax
\usepackage{textcomp}
\usepackage{amsthm}
\usepackage{amssymb}  
\usepackage{mathtools}
\usepackage[utf8]{inputenc}
\usepackage{comment}
\usepackage{algorithm,algpseudocode}
\algblock{Input}{EndInput}
\algnotext{EndInput}
\algblock{Output}{EndOutput}
\algnotext{EndOutput}

\usepackage{breqn}
\usepackage{graphicx}
\usepackage[version=4]{mhchem}
\usepackage{siunitx}
\usepackage{longtable,tabularx}
\usepackage{graphics} 
\usepackage{float}
\usepackage{epsfig} 
\usepackage{times} 
\usepackage{color}
\usepackage[dvipsnames]{xcolor}
\usepackage{pifont}
\usepackage{enumitem}
\usepackage{bm}
\usepackage{bbm}
\usepackage{soul}
\makeatletter
\newcommand{\multiline}[1]{%
  \begin{tabularx}{\dimexpr\linewidth-\ALG@thistlm}[t]{@{}X@{}}
    #1
  \end{tabularx}
}

\usepackage{arydshln}

\makeatother
\usepackage{hyperref}
\hypersetup{
    colorlinks=true,
   linkcolor=black,
    filecolor=blue,      
   urlcolor=blue,
    pdftitle={Overleaf Example},
    pdfpagemode=FullScreen,
    }

\newcommand{\R}{\mathbb{R}}

\newtheorem{problem}{Problem}

\DeclareMathOperator*{\argmin}{arg\,min}

\usepackage{todonotes}

\newcounter {dagger} 
\begin{document}
\title{Floodgates up to contain the DeePC and limit extrapolation}

\author{Mohammad S. Ramadan,\,Evan Toler,\,Mihai Anitescu
\thanks{The authors are with the Mathematics and Computer Science Division, Argonne National Laboratory, Lemont, IL 60439, USA,  {\tt\footnotesize mramadan@anl.gov, etoler@anl.gov, anitescu@mcs.anl.gov.}}}

\maketitle
\thispagestyle{empty}
\pagestyle{empty}

\begin{abstract}
Behavioral data-enabled control approaches typically assume data-generating systems of linear dynamics. This may result in false generalization if the newly designed closed-loop system results in input-output distributional shifts beyond learning data. These shifts may compromise safety by activating harmful nonlinearities in the data-generating system not experienced previously in the data and/or not captured by the linearity assumption inherent in these approaches. This paper proposes an approach to slow down the distributional shifts and therefore enhance the safety of the data-enabled methods. This is achieved by introducing quadratic regularization terms to the data-enabled predictive control formulations. Slowing down the distributional shifts comes at the expense of slowing down the exploration, in a trade-off resembling the exploration vs exploitation balance in machine learning.
\end{abstract}

\begin{keywords}
Data-enabled control, data-driven control, adaptive control, DeePC algorithm.
\end{keywords}
\section{Introduction} \label{section: Introduction}
In the 1980s, Charles E. Rohrs, a PhD student then, challenged the adaptive control community and sought to invalidate their mainstream premises and results \cite{anderson2005failures,rohrs1985robustness,aastrom1983analysis}. Rohrs argued that almost every physical system has unmodeled (parasitic) high-frequency dynamics which can be unstable or harmful. Given that adaptive control algorithms at that time led to highly nonlinear loops even for simple linear systems, Rohrs emphasized that they could internally generate high-frequency signals, such that these unmodeled dynamics can be excited, possibly leading to failures and instability. In this paper we present an analogous argument addressing the modern data-driven control algorithms. Instead of Rohrs' unmodeled high frequency dynamics (which does not apply for these approaches if the data matrices are not adjusted in a high rate), we target (parasitic) unmodeled nonlinear dynamics, which also exist in almost every physical system. Our proposed adjustment is a simple quadratic regularization term that does not carry a significant additional computational or algorithmic burden.

Data-driven control, the behavioral approach in particular, presents itself as a shift from conventional control design paradigms that rely on explicitly modeling data-generating systems. The behavioral approach, pioneered by Willems \cite{willems2005note} and further developed in subsequent research, describes a dynamic system through a set of its trajectories, skipping the need for a model, e.g. a state-space or a transfer function \cite{maupong2017data,markovsky2021behavioral}. This perspective in inferring system behavior directly from data relies on the Fundamental Lemma \cite{willems2005note}, which establishes conditions under which data collected from persistently exciting inputs spans the system's behavior space. However, limitations of this approach lie not only in its inherent assumption of linearity - since the underlying theory is built for linear time-invariant (LTI) systems, but also in the degree to which the LTI assumption can be forgotten from the control designer perspective. Having explicit LTI state-space models or transfer functions carry this assumption in themselves, where in the behavioral formulation this is less evident. Whether model-based or model-free, the LTI assumption allows the control design algorithm to anticipate a universal generalization of linearity across all future trajectories, which can introduce challenges when the underlying data-generating systems in consideration exhibit destabilizing/harmful nonlinearities not captured previously by data. Therefore, forgetting or not accounting for this assumption can be of serious consequences.

To prevent the consequential universal generalization of the LTI assumption, we propose convex regularization terms that can be added to the general formulation of data-driven predictive control methods both for the model-free and model-based approaches. These regularization terms are built on the Mahalanobis distance between the future closed-loop trajectories and the past data distribution. The Mahalanobis distance is the multivariate generalization to the number of standard deviations from the mean in the univariate Gaussian case. It can be used to detect outliers of a distribution, hence for our purpose, it will be used to penalize closed-loop trajectories seen as outlier with respect to the data distribution, limiting extrapolating the LTI assumption beyond data.

Enforcing  consistency with past data has been previously investigated under the unfalsified control paradigm \cite{cabral2004unfalsified}, which models consistency as a Boolean value: data consistent or not. Also, offline RL \cite{agarwal2020optimistic} enforces this consistency with the motivation of data efficiency by initially limiting exploration. Offline RL algorithms are generally complex \cite{fujimoto2021minimalist}; since they rely on stochastic nonlinear programming methods. Instead, in previous works \cite{ramadan2024data,ramadan2024dampening}, we developed computationally efficient methods that blend with modern control design techniques and result in convex semi-definite programs. In this paper, we extend our work to enforce data consistency for the data-driven control approaches and the data-enabled predictive control (DeePC) algorithm \cite{coulson2019data}. We call this extension the floodgates up DeePC; as a metaphor to its function in preventing premature generalization which may result in drastic distributional shifts (or flooding beyond the learning data).

We present numerical simulations of a simple yet telling example showing what can go wrong when the DeePC falsely extrapolates the linearity assumption over an underlying nonlinear data-generating system.

\section{Data-conforming data-enabled predictive control} \label{section:the predictive case}
In previous papers, we presented formulations to enforce consistency to learning data for the unconstrained infinite-horizon formulation \cite{ramadan2024data}, and for the case of multiplicative uncertainties accounting for unknown parameters and gain-scheduling \cite{ramadan2024dampening}. However, in many situations, a data-driven control design objective may involve the satisfaction of input and state/output constraints. For this purpose, predictive control is a natural control design framework due to its finite receding-horizon formulation \cite{morari1999model}.

In this section we present an ``indirect'' (model-based) data-driven model predictive control (MPC) and a ``direct'' (directly from data, hence model-free) data-enabled predictive control (DeePC) algorithm \cite{coulson2019data}, both modified to account for data consistency. The proposed modifications are of no additional significant algorithmic or computational complexity to the basic formulations of these methods.

\subsection{The Mahalanobis distance}
The Mahalanobis distance can be used to represent a confidence region (or set), a multidimensional generalization of confidence intervals in the univariate case. It can also be used to detect outliers with respect to Gaussian distributions. This will be our use of this distance: to penalize such outliers of the new design compared to the learning data distribution.

The Mahalanobis distance $d_M$, of a point $x\in\R^{r}$ with respect to a known multivariate Gaussian on that space $\mathcal{N}_0= \mathcal{N}(\mu_0, \Sigma_0)$ with $\Sigma_0 \succ 0$, is given by
\begin{align*}
    d_M(x; \mathcal{N}_0) = \sqrt{(x-\mu_0)^\top \Sigma_0^{-1} (x-\mu_0)}.
\end{align*}
This distance is related to confidence intervals of univariate Gaussian densities. In the univariate case, the $2\sigma$ set $\{x\in \R \mid d_M \leq 2\}$ corresponds to the (approximately) $95\%$ confidence interval. In the multivariate case, say, on $\R^r$, we can write $d_M^2(x; \mathcal{N}_0) = z(x)^\top z(x)$, where the random vector $z(x)=\Sigma_0^{-1/2}(x-\mu_0)$ consists of $r$ independent entries, each of which is an independent standardized Gaussian random variable (since $x$ is sampled according to $\mathcal{N}_0$). Therefore, $d_M^2(x; \mathcal{N}_0) = z(x)^\top z(x)$ is a random variable sampled according to the $r-$ degrees of freedom chi-square density $\chi^2(r)$. This makes it possible to relate the sets $\{x\in \R^r \mid d_M \leq d_0\}$, for any $d_0 \geq 0$, to their confidence level, given by the probability of the corresponding set (in $z$) under $\chi^2(r)$.

For $\delta>0$, the $(1-\delta)100\%$ confidence set can be found by solving for $d^\star$ (a root-finding problem of a monotonic function)
\begin{align}
    1-\delta = \mathcal P\left (\frac{r}{2}, \frac{d^\star}{2} \right), \label{eq:chi-squared cdf}
\end{align}
where $\mathcal P$ is the regularized gamma function and the cumulative distribution function of the degree$-r$ chi-squared density $\chi^2(r)$. The $(1-\delta)100\%$ confidence set is then defined by
\begin{align*}
    \{ x \in \R^r \mid d_M^2(x;  \mathcal{N}_0) \leq d^\star \}, 
\end{align*}
which represents a hyperellipsoid. Importantly from an optimization viewpoint, this is a convex set, which will admit fast and efficient numerical methods when used as an optimization feasibility constraint or included in the construction of a regularization term.

\subsection{Data-conforming data-driven MPC}

Consider the dynamic system
\begin{align}
x_{k+1}&=f(x_k, u_k, w_k), \label{eq:stateEquation}
\end{align}
where $x_k\in\mathbb R^{r_x}$ is the state and $u_k\in\mathbb R^{r_u}$ is the control input, both are possibly required to be contained in the linear constraint sets $\mathbb X$ and $\mathbb U$, respectively. The exogenous disturbance $w_k\in\mathbb R^{r_w}$ is independent and identically distributed. The function $f$ is bounded but unknown and possibly nonlinear. However, previous state-input data are available and carry the persistence of excitation assumption \cite[condition~(6)]{de2019formulas}. That is, given a natural number $T \gg (r_u +1) r_x + r_u$, the control realization $\{u_{k}\}_{k=0}^{T}$ and the corresponding state realization $\{x_{k}\}_{k=0}^{T}$ in the past data produce the data matrix
\begin{equation} \label{eq:data matrix}
     {D} := 
    \begin{bmatrix}
          X\\
          U
    \end{bmatrix}
    =
    \begin{bmatrix}
        x_0 & \hdots &x_{T}\\
        u_0 & \hdots &u_{T}
    \end{bmatrix}
\end{equation}
with full row rank $r_x + r_u$.

It is required to design a data-driven predictive controller minimizing the finite-horizon cost
\begin{align}
\begin{aligned}
    J^N(x_0,\{u_k\}_{k}) &= x_N^\top Q x_N +
    \sum_{k=0}^{N-1} x_k^\top Q x_k + u_k^\top R u_k, \label{eq:costFunction MPC}
\end{aligned}
\end{align}
where $Q \succeq 0$ and $R \succ 0$ are positive semi-definite and positive definite, respectively. It is also required to satisfy the constraints $x_k \in \mathbb X$, $u_k \in \mathbb U$, and $x_N \in \mathbb X_N$, where $\mathbb X$, $\mathbb U$ and $\mathbb X_N$ are linear (polyhedral) sets. The terminal constraint set $\mathbb X_N \subset \mathbb X$ is possibly engineered to achieve some recursive feasibility guarantees.

We restrict the control policy to have the form of an open- and linear closed-loop components: $u_k = K x_k + \tilde u_k$. 

Given the empirical data distribution is captured by\footnote{The Gaussianity assumption might be strong but it can be implied by a more relaxed condition than the small signal model. For the small signal model assumption, the model is assumed to behave according to a linear model as long as the input signal is small. On the other hand, one way to imply the Gaussianity assumption is to assume, given some initial state distribution, that the model behaves according to a linear model under a certain control law, but may act according to a different linear model under a different control law.} $ \mathcal{N}_{data} =  \mathcal{N}(\mu_{data}, \Gamma_{data})$, with $\Gamma_{data}$ representing the empirical state-input joint covariance $\Gamma_{data} = (T+1)^{-1}\left [D -\mu_{data}\right] \left [D -\mu_{data}\right]^\top$, some regularization terms we may use to enforce conforming to data are
\begin{align}
    F(x_k, u_k)&=d_M^2([x_k^\top, u_k^\top]^\top;  \mathcal{N}_{data}), \label{eq:first reg}\\
    &= \left( 
    \begin{pmatrix}
        x_k\\
        u_k
    \end{pmatrix}
    -\mu_{data}\right)^\top
    \Gamma_{data}^{-1}
    \left( 
    \begin{pmatrix}
        x_k\\
        u_k
    \end{pmatrix}
    -\mu_{data}\right)
    \nonumber\\
    &\hskip-5mm \text{or,}\nonumber\\
    F(x_k, u_k)&= \max \Big \{0,  d_M^2([x_k^\top, u_k^\top]^\top;  \mathcal{N}_{data}) - d^\star \Big \}\label{eq:second reg},
\end{align}
where the value for $d^\star$ is chosen by the control designer. The second term adds no penalty on the controlled trajectories as long as they fall within the confidence set defined by $d^\star$. In this context, we distinguish between confidence and likelihood: a large value $d^\star$ in \eqref{eq:second reg} corresponds to a huge confidence set ($\approx 100\%$ confidence, or the whole space), hence, rendering \eqref{eq:second reg} meaningless. In other words, no controlled trajectory can be an outlier to the learning distribution, and consistency is not enforced. Therefore, a confidence region is to be selected based on its likelihood content, such that it penalizes sensible outliers.

\begin{problem}\label{prob:Data-conforming MPC}
    Data-conforming data-driven MPC\footnote{$X_0, U_0$ are the first $N$ columns of $X$ and $U$, while $X_1$ is the last $N$ columns of $X$.}:
\begin{align*}
        &\min_{\{\tilde u_k\}_{k=0}^{N-1},\{x_k\}_{k=1}^{N}} J^N(x_0,\{u_k\}_{k=0}^{N-1}) + \gamma\sum_{k=0}^{N-1} F(x_k, u_k) \\
        \text{s.t. } &x_{k+1} = \left [\widehat A + \widehat B K \right ] x_k + \widehat B u_k,\,k=0,\hdots,N-1,\\
        &u_k \in \mathbb U,\, x_k \in \mathbb X,\,k=1,\hdots,N-1,\, x_N \in \mathbb X_N,\\ 
        & [\widehat A,\widehat B] =\argmin_{ A, B} \left \lVert   X_1 - [A,  B] 
        \begin{bmatrix}
              X_0 \\
              U_0
        \end{bmatrix}
        \right \rVert_F,\\
        &u_k = K x_k + \tilde u_k, \text{ $K$ stabilizes $(\widehat A,\widehat B).$}
\end{align*}
\end{problem}

Problem~\ref{prob:Data-conforming MPC} is solved at each time step via identifying the model $(\widehat A,\widehat B)$, designing $K$ through\footnote{The control gain $K$ and the terminal constraint set $\mathbb X_N$ can be designed under robustness conditions for multiplicative uncertainties \cite{fleming2014robust}, since the identified model is uncertain, in order to achieve some robust or stochastic recursive feasibility guarantees.}, say, an unconstrained infinite-horizon LQR, then solving the optimization problem and applying $u_0$ at the current time step.

Problem~\ref{prob:Data-conforming MPC} assumes a one-time experiment has been done and data has been collected, and therefore it is not designed to be adaptive and/or admit exploration automatically. To allow exploration and change in parameters, a control designer can lower the value of $\gamma$, which serves as the hyperparameter controlling exploration vs exploitation balance, while a persistently exciting input is injected to keep the future data sufficiently informative for iterative learning. In the MPC context, a control designer needs to incorporate special constraints \cite{marafioti2014persistently} to guarantee persistence of the input signal.

\subsection{Data-conforming DeePC}
The DeePC algorithm \cite{coulson2019data} is a direct data-driven optimal control strategy that computes an output feedback control for unknown systems. In the output regulation problem, that is, the state $x_k$ in \eqref{eq:stateEquation} is not observed directly, but partially through some output signal $y_k \in \R^{r_y}$, 
\begin{align}
    y_k = g(x_k,\nu_k), \label{eq:outputEquation}
\end{align}
where $\nu_k \in \R^{\nu}$ is independent and identically distributed. The DeePC is described by the following problem.

\begin{problem}\label{prob:Standard DeePC}
Standard DeePC:
\begin{align*}
         \min_{g,u,y,\sigma}& y_N^\top Q_y y_N + \sum_{k=0}^{N-1} y_k^\top Q_y y_k + u_k^\top R u_k + \\
         &\hskip15mm \lambda_g \lVert g \rVert_1 + \lambda_\rho \lVert \rho \rVert_1 \\
         &\text{s.t.}
         \begin{bmatrix}
            {\color{black}U_p} \\
            {\color{black}Y_p} \\
            {\color{black}U_f} \\
            {\color{black}Y_f}
        \end{bmatrix}
        g
        =
        \begin{pmatrix}
            {\color{black}u_{\text{ini}}} \\
            {\color{black}y_{\text{ini}}} + \rho \\
            {\color{black}u} \\
            {\color{black}y}
        \end{pmatrix},\quad u_k \in \mathbb U, y_k \in \mathbb Y,
\end{align*}
\end{problem}
\noindent
and the constraint sets $\mathbb U$ and $\mathbb Y$ are assumed polyhedral. The parameters $\lambda_g,\lambda_\rho$ are regularization weights. The slack vector $\rho$ guarantees feasibility under output noise or possible existence of nonlinearities (for more discussion on the function of $\lambda_g,\lambda_\rho, \rho$ and how to tune them, please consult \cite{coulson2019data}). The data matrices
\begin{align*}
    \underset{(T_{\text{ini}}+N)r_u \times (T - (T_{\text{ini}}+N) + 1)}{\begin{bmatrix}
        U_{P}\\
        U_f
    \end{bmatrix}}
     :=  {H}_{T_{\text{ini}}+N} \left (U^d \right ),\\
     \underset{(T_{\text{ini}}+N)r_y \times (T - (T_{\text{ini}}+N) + 1)}{\begin{bmatrix}
        Y_{P}\\
        Y_f
    \end{bmatrix}}
     :=  {H}_{T_{\text{ini}}+N} \left (Y^d \right ),
\end{align*}
where $T_{\text{ini}} \geq (r_u+1)r_x + r_u$ is a positive integer and $U_P,\,Y_P$ are the first $T_{\text{ini}}r_u$ and $T_{\text{ini}}r_y$ rows of ${H}_{T_{\text{ini}}+N} \left (U^d \right )$ and ${H}_{T_{\text{ini}}+N} \left (Y^d \right )$, all respectively. The data matrices $U^d := \left [u^d_0 \hdots u^d_T \right ]$ and $Y^d := \left [y^d_0 \hdots y^d_T \right ]$ are persistently excited in the sense defined in \cite{coulson2019data}, and the Hankel matrices above, for a positive integer $l$ and vectors $a_i \in \R^{r_a}$, are defined according to
\begin{align}
     \underset{l r_a \times (m-l+1)}{{H}_{l} \left (
    \begin{bmatrix}
        a_0 & a_1& \hdots & a_{m}
    \end{bmatrix}
    \right )}
    = 
    \begin{bmatrix}
    a_0 & a_1 & \hdots & a_{m - l}\\
    a_1 & a_2 & \hdots & a_{m - l + 1}\\
    \vdots & &\ddots & \vdots \\
    a_l & a_{l+1} & \hdots & a_m
    \end{bmatrix}.
\end{align}
 The weighting matrices $Q_y \succeq 0,\, R \succ 0$, and the vectors
\begin{align}
    u_{\text{ini}}
    &:= \text{col}(u_{-{T_{\text{ini}}}}, \hdots, u_{-1}) := 
    \begin{pmatrix}
    u_{-{T_{\text{ini}}}}\\
    \hdots \\
    u_{-1}
    \end{pmatrix},
    \\
     y_{\text{ini}}
    &:=
    \text{col}(y_{-{T_{\text{ini}}}}, \hdots, y_{-1}), \label{eq:uy ini data}
\end{align}
denote the data points immediately before the current time ($k=0$) and together are used as the surrogate description of the initial condition of the system. The input-output decision variables are
\begin{align}
    u &:= \text{col}(u_{0}, \hdots, u_{N-1}),\\
     y &:= \text{col}(y_{0}, \hdots, y_{N}).
\end{align}

Since the output sequence $y_k$ is not necessarily a state of the system \eqref{eq:stateEquation} (possibly non-Markov), a repeated occurrence of a pair $(u_k,y_k)$ in different occasions does not imply that the state of the underlying system is the same in these occasions. Hence, we have two options when applying the Mahalanobis distance: (i) approximate a state-space model, say, by an eigensystem realization algorithm \cite{peterson1995efficient}, and then track the state via some state observer or (ii) use the columns of the input/output Hankel matrices of past data as surrogate descriptions to the state. We pick the latter for no reason other than keeping the DeePC in its direct (model-free) form.

Let
\begin{align}
    \mathcal H_{data} &:=
    \frac{1}{T+1-T_{\text{ini}}} \times \nonumber\\
    &\hskip-5mm \begin{bmatrix}
         U_p - \mu_{data}^u \\
         Y_p - \mu_{data}^y
    \end{bmatrix}
    \begin{bmatrix}
         U_p - \mu_{data}^u\\
         Y_p - \mu_{data}^y
    \end{bmatrix} ^ \top \succeq 0.
\end{align}
Since if $T_{\text{ini}}$ is larger or equal to the order of the underlying system, by Lemma~4.1 in \cite{coulson2019data}, the past state sequence of a linear model of such order can be retrieved from the above Hankel matrices. That is, the columns of the above Hankel matrices can serve as the surrogate (not necessarily minimal) state. The vectors $\mu_{data}^u$ and $\mu_{data}^y$ are the averages of the columns of $U_p$ and $Y_p$, respectively. $\mathcal H_{data}$  is not guaranteed to be invertible, so we instead use
\begin{align}
    \mathcal H_{data}^\varepsilon = \mathcal H_{data} + \varepsilon \mathbb{I}, \label{eq:uy cov with epsilon}
\end{align}
which is positive definite for $\varepsilon>0$ and hence invertible. The corresponding density is given by
\begin{align*}
    \mathcal{N}_{\mathcal{H}}=\mathcal{N}(\mu_{data}^{uy}, \mathcal H_{data}^\varepsilon ),
\end{align*}
where $\mu_{data}^{uy}=\text{col} (\mu_{data}^{u},\, \mu_{data}^{y})=\left [(\mu_{data}^{u})^\top, ( \mu_{data}^{y})^\top \right]^\top$.

Using \eqref{eq:uy cov with epsilon}, we construct analogous terms to \eqref{eq:first reg} or \eqref{eq:second reg},
\begin{align}
    F_{\mathcal H}(\Psi_k)&=d_M^2(\Psi_k;  \mathcal{N}_{\mathcal H}), \label{eq:first reg hankel}\\
    &\hskip-5mm \text{or,}\nonumber\\
    F_{\mathcal H}(x_k, u_k)&= \max \Big \{0,  d_M^2(\Psi_k;  \mathcal{N}_{\mathcal{H}}) - d^\star_{\mathcal{H}} \Big \}\label{eq:second reg hankel},
\end{align}
where the $(r_u + r_y)T_{\text{ini}} \times 1$ vector $\Psi_k = \text{col}(u_{k-T_{\text{ini}}}, \hdots, u_k, y_{k-T_{\text{ini}}}, \hdots, y_k)$, $k=0,\hdots,N-1,$ with the $u_k$s and $y_k$s of negative subscripts (recent past data) are given by \eqref{eq:uy ini data}, and $d^\star_{\mathcal{H}}$ can be found similar to $d^\star$ according to the commulative distribution function \eqref{eq:chi-squared cdf}.

The Mahalanobis distance-based data-conforming DeePC is as follows.

\begin{problem}\label{prob:Data-conforming DeePC}
Floodgates up DeePC:
\begin{align*}
         &\min_{g,u,y,\sigma} y_N^\top Q_y y_N + \sum_{k=0}^{N-1} \left \{ y_k^\top Q_y y_k + u_k^\top R u_k + \gamma F_{\mathcal H}(\Psi_k) \right \} \\
         &\hskip 30mm \lambda_g \lVert g \rVert_1 + \lambda_\rho \lVert \rho \rVert_1, \\
         &\text{s.t.}
         \begin{bmatrix}
            {\color{black}U_p} \\
            {\color{black}Y_p} \\
            {\color{black}U_f} \\
            {\color{black}Y_f}
        \end{bmatrix}
        g
        =
        \begin{pmatrix}
            {\color{black}u_{\text{ini}}} \\
            {\color{black}y_{\text{ini}}} + \rho \\
            {\color{black}u} \\
            {\color{black}y}
        \end{pmatrix}, \quad u_k \in \mathbb U, y_k \in \mathbb Y,
\end{align*}
\end{problem}

Problem~\ref{prob:Data-conforming DeePC} is similar to Problem~\ref{prob:Standard DeePC} but with the Mahalanobis distance regularization term. Problem~\ref{prob:Data-conforming DeePC} is still a quadratic program if $\mathbb U$ and $\mathbb Y$ can be described by linear inequalities. The hyperparameter $\gamma$, similar to the data-conforming data-driven control, can be used to control the exploration vs exploitation balance.

\section{Numerical simulations} \label{section: Numerical}
The scalability of our proposed data-conforming formulation of the DeePC is self-evident, as it preserves the nature of the DeePC as a quadratic program. Therefore, in this section, we instead choose to work with a low order model to illustrate the motivation and effectiveness of our approach, and what can wrong with the regular DeePC algorithm.

Suppose the data-generating system described by \eqref{eq:stateEquation} and \eqref{eq:outputEquation} has the form\footnote{The results of this section can be reproduced using our open-source \textsc{Julia} code found at \href{https://github.com/msramada/floodGatesUp-DeePC}{https://github.com/msramada/floodGatesUp-DeePC}.}
\begin{equation} \label{eq:example state-space equation}
    \begin{aligned}
            \begin{pmatrix}
                x_{1,k+1}\\
                x_{2,k+1}
            \end{pmatrix}=
            x_{k+1} = &
    \begin{bmatrix}
        .98& .1\\
        0& .95
    \end{bmatrix}x_k + 
    \begin{pmatrix}
        \theta x_{2,k}^2\\
        0
    \end{pmatrix}
    + \\
    &\begin{bmatrix}
        0\\
        0.1 + \theta \tanh{x_{1,k}}
    \end{bmatrix}u_k +
    w_k,\\
    y_k = &
    \begin{bmatrix}
        0 & 1
    \end{bmatrix} x_k + \nu_k,
    \end{aligned}
\end{equation}
with $\theta = 1/9$ and the noises $w_k,\nu_k$ are Gaussian with zero means and covariances $\text{diag}(0.1,\,0.05)$ and $0.1$, respectively. Toward testing our proposed approach, Problem~\ref{prob:Data-conforming DeePC}, and test it against the regular DeePC, Problem~\ref{prob:Standard DeePC}, we first conduct a simulation of $T=200$ time-steps and collect the input-output pairs (together with the state $x_k$, which is assumed hidden and will not be used in any algorithm, but will be used only for demonstration purposes). In this simulation we use an output feedback control $u_k = K_0 y_k$, where $K_0=\left [ -6.0\right ]$. The output noise $\nu_k$ guarantees the persistence of excitation condition of the input. We use $T_{\text{ini}}=4$, $N=8$, $\lambda_g = \lambda_\rho = 1$. For both problems, Problem~\ref{prob:Standard DeePC} and Problem~\ref{prob:Data-conforming DeePC}, we use $Q_y=1,R=2$, with no input or output constraints, and for Problem~\ref{prob:Data-conforming DeePC}, we adopt the regularization term \eqref{eq:first reg hankel}. We then run these two problems in feedback, starting from a zero initial condition, for $T_{sim}=200$ time-steps and record their corresponding closed-loop data.

\begin{figure}
\centering 
\includegraphics[width=2.8in,height=2.8in]{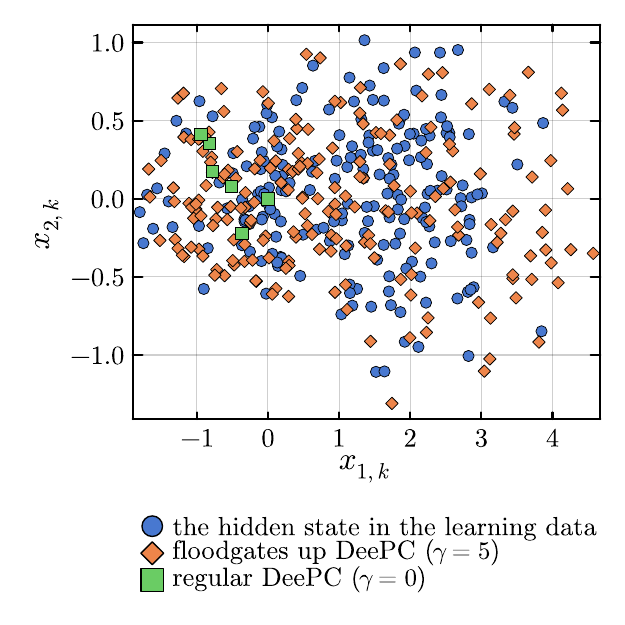} 
\caption{X-axis and y-axis correspond to the state space, $x_{1,k}$ and $x_{2,k}$, respectively. The blue circles are the (hidden) state data corresponding to the initial simulation experiment, using $u_k=K_0 y_k$. The orange diamonds are the state data resulting from solving Problem~\ref{prob:Data-conforming DeePC} with $\gamma=5$, showing a similarity in state distribution to that of the collected data. The green squares are the states resulting from using the regular DeePC algorithm, Problem~\ref{prob:Standard DeePC} (or, equivalently, Problem~\ref{prob:Data-conforming DeePC} with $\gamma=0$), and only few of these squares are shown as the system has gone unstable after few time-steps.}\label{fig:state distributions}
\end{figure}

Figure~\ref{fig:state distributions} shows the state sequences of the collected data and the closed-loop systems resulting from Problems~\ref{prob:Standard DeePC} and \ref{prob:Data-conforming DeePC}. These state sequences are hidden, not accessed by any algorithm, and only recorded in simulation for illustration. Notice that the control resulting from Problem~\ref{prob:Data-conforming DeePC} results in a state distribution resembling that of the learning data, and hence avoids the nonlinearity $\theta x^2_{2,k}$, which was not yet significant in the learning data. On the other hand, the regular DeePC, with its linearity assumption, can lead to a state sequence beyond the distribution of the learning data, to regions where not yet discovered nonlinearities are significant, and hence is unstable, which is the reason only few time-steps are shown in Figure~\ref{fig:state distributions} before blowing up.

Due to the stochastic nature of this control problem, we repeat the above procedure for a hundred times, starting with data collection of $T=200$ input-output points, then run Problems~\ref{prob:Standard DeePC} and \ref{prob:Data-conforming DeePC} in closed-loop, with zero initial conditions, and for $T_{sim}=100$ time-steps each. We count a simulation as unstable if $|y_k|>50$ for any $k=0,\hdots,T_{sim}$. As a result,
\begin{itemize}
    \item the regular DeePC algorithm, Problem~\ref{prob:Standard DeePC}, led to instability $100\%$ of the times,
    \item while Problem~\ref{prob:Data-conforming DeePC} resulted in instability only $2\%$ of these runs.
\end{itemize}
This is expected; as Problem~\ref{prob:Data-conforming DeePC} avoids premature extrapolation beyond seen data, and hence avoids harmful nonlinearities that were not significant in the learning data.

It is legitimate to question a control algorithm that generates the data distribution similar to that of the conducted experiment. Indeed, control algorithms are designed to achieve some objective that might differ from the behavior seen in the learning experiment. Our argument is that immediately seeking this different objective might result in dangerous distributional shifts. Therefore, it is of the control designer duty to start with a high value of $\gamma$, dampening any distributional shifts at first, then gradually reduce it to allow some exploration while being observant and cautious of any unsafe behavior. The adaptive control literature \cite{astrom1985commentary,anderson2005failures} recommends to ``model as much as possible'', and our approach can provide the safety and the time necessary for the control practitioner to model and interpret any activation of a nonlinearity.

\section{Conclusion} \label{section: Conclusion}
In this paper we address a problem inherent in modern data-driven and data-enabled control methods, namely, the problem of the premature and possibly false extrapolation beyond the learning data. In particular, we equip the DeePC algorithm with a regularizing term that enforces consistency to the learning data, preventing any sudden distributional shifts after its implementation in closed-loop. Due to the quadratic form of this regularization term, it does not alter the DeePC algorithm as a quadratic program, preserving its computational efficiency and scalability. Further research is being pursued to equip data-driven control approaches with new safety measures and dampen their distributional shifts in the hidden state-space.
\bibliographystyle{IEEEtranS}
\bibliography{References}

\vspace{0.1cm}

\begin{flushright}
	\scriptsize \framebox{\parbox{2.5in}{Government License: The
			submitted manuscript has been created by UChicago Argonne,
			LLC, Operator of Argonne National Laboratory (``Argonne").
			Argonne, a U.S. Department of Energy Office of Science
			laboratory, is operated under Contract
			No. DE-AC02-06CH11357.  The U.S. Government retains for
			itself, and others acting on its behalf, a paid-up
			nonexclusive, irrevocable worldwide license in said
			article to reproduce, prepare derivative works, distribute
			copies to the public, and perform publicly and display
			publicly, by or on behalf of the Government. The Department of Energy will provide public access to these results of federally sponsored research in accordance with the DOE Public Access Plan. http://energy.gov/downloads/doe-public-access-plan. }}
	\normalsize
\end{flushright}	
\end{document}